\documentstyle[12pt]{article}
\catcode`\@=11
\def\eqnarray{\let\@currentlabel=\theequation\refstepcounter{equation}
    \global\@eqnswtrue
    \global\@eqcnt\z@\tabskip\@centering\let\\=\@eqncr
    $$\halign to \displaywidth\bgroup\@eqnsel\hskip\@centering
      $\displaystyle\tabskip\z@{##}$&\global\@eqcnt\@ne 
       \hfil${{}##{}}$\hfil
      &\global\@eqcnt\tw@ $\displaystyle\tabskip\z@{##}$\hfil 
       \tabskip\@centering&\llap{##}\tabskip\z@\cr}
\def\lefteqn#1{\hbox to 4\arraycolsep{$\displaystyle #1$\hss}}
\newcommand{\beq}{\begin{equation}}
\newcommand{\eeq}{\end{equation}}
\def\IR{{\hbox{{\rm I}\kern-.2em\hbox{\rm R}}}}

\begin{document}

\begin{center}
{\bf
DILATONIC DARK MATTER AND UNIFIED COSMOLOGY \\
         -- A NEW PARADIGM --}
\vskip 1cm
Y.\ M.\ Cho\footnote{e-mail: yongmin@ymcho.snu.ac.kr}\vskip 1cm
{\it
Asia Pacific Center for Theoretical Physics, \\ 
and \\
Department of Physics, College of Natural Sciences \\
Seoul National University, Seoul 151-742, Korea } \\

\vskip 1cm
Y.\ Y.\  Keum\footnote{e-mail: keum@apctp.kaist.ac.kr}\vskip 1cm
{\it Asia Pacific Center for Theoretical Physics \\
 207-43 Cheongryangri-Dong Dongdaemun-Gu, Seoul 130-012, Korea }

\end{center}
\vskip 1cm
\noindent {\bf Abstract}
\vskip 0.7cm
We study the possibility that the dilaton --
the fundamental scalar field which exists in all the existing
unified field theories -- plays 
the role of the dark matter of the
universe. We find that the condition
for the dilaton to be the dark matter strongly restricts its 
mass to be around 0.5 keV or 270 MeV. 
For the other mass ranges, the dilaton
either undercloses or overcloses the universe. 
The 0.5 keV dilaton has the 
free-streaming distance of about 1.4 Mpc and becomes 
an excellent candidate of a warm dark matter, while the 270 MeV one
has the free-streaming distance of about 7.4 pc 
and becomes a cold dark matter.
We discuss the possible ways to detect the dilaton experimentally.

\vskip 1cm
\noindent {\bf I. Introduction}
\vskip 0.7cm

The standard big bang cosmology has been very successful in 
many ways. For example it naturally explains the
Hubble expansion, the cosmic microwave background, and the primordial
nucleosynthesis of the universe.
But at a deeper level the model also raises more challenges.
Is the inflation really necessary, and if so, what drives the inflation? 
What kind of dark matter, and how much of it,
does the universe contain?
How was the observed structure of the universe formed,
and how did the dark matter affect this?
There may be two attitudes that one can adopt in dealing with
these challenges.
One is a phenomenological approach.
Here one introduces a minimum number of new parameters (a cosmological 
constant, for example) to the standard model,
and try to obtain  a best model
which can fit as many data as possible.
This is indeed the popular approach at present.
The other one is a theoretical approach.
Here one tries to construct a model which
is most appealing from the logical point of view, based
on the fundamental principles. 
For example, if one believes in the unification
of all interactions, one may  
ask what is the best cosmological model
that one can obtain from the unified theory. 
In this paper we will take the second attitude 
and discuss how the unification of all interactions could modify the 
big bang cosmology. The reason for this is 
partly because few people
in cosmology takes this attitude, 
and partly because  the popular approach has been
thoroughly discussed by many authors. We believe that our approach
could add a new insight to the cosmology.

The standard big bang cosmology may need a generalization from
the observational point of view.
But it must be emphasized that there is also a strong 
motivation to generalize it from the theoretical point of view.
This is because the standard model is based on the Einstein's
theory of gravitation, which itself may need a generalization.
Of course the Einstein's theory has been a
most beautiful and successful theory of gravitation.
But from the logical point of view
there are many indications that something 
is missing in the Einstein's theory. 
We mention just a few:

\noindent (1) The unification of all interactions may require a
generalization of Einstein's theory. 
All the known interactions are mediated by the spin-one or
spin-two fields.
However, the unification of all interactions
inevitably requires the existence of a fundamental spin-zero field.
In fact all modern unified theories---the Kaluza-Klein theory,
the supergravity, and the superstring--- contain a fundamental scalar
field. What makes this scalar field unique is that, 
unlike others  like the Higgs field, 
it couples directly to the (trace of) energy-momentum tensor of the matter
field.
As such it should generate a new force which will 
modify the Einstein's gravitation  
in a fundamental way.

\noindent (2) The Newton's constant $G$, which is supposed to be one of the
fundamental constants of Nature, plays a crucial role in Einstein's
theory. But the ratio between the electromagnetic fine structure 
constant $\alpha_{e}$ and the gravitational fine structure constant
$\alpha_g$ of the hydrogen atom is too small to be considered natural,
\begin{eqnarray}
\alpha_g / \alpha_{e} \simeq 10^{-40}.
\end{eqnarray}
This implies that not both of them may be regarded as the fundamental
constants of Nature. 
From this Dirac\cite{dirac} conjectured that $G$ may not be a fundamental
constant, but in fact a time-dependent parameter. If so,
one must treat it as a fundamental scalar field which couples
to all matter fields.
Obviously the Dirac's conjecture requires a drastic generalization
of Einstein's theory.

\noindent (3) The conformal transformation which changes
the scale of the space-time metric at each space-time point,
\begin{eqnarray}
g_{\mu\nu} \longrightarrow e^{\sigma(x)}g_{\mu\nu} ,
\end{eqnarray}
is not a fundamental symmetry of Nature.
At a deeper level (in the high energy limit), however, there is a real possibility that
the conformal invariance and its scale factor may play
an important role in physics.
But in Einstein's theory there is no place where the conformal invariance and
 its 
scale factor could play a role.

\noindent (4) Finally, in cosmology the inflation at the early
stage of evolution may be unavoidable.
But for a successful inflation we need a dynamical mechanism
which can (not only initiate but also) stop
 the inflation smoothly.
Unfortunately the Einstein's gravitation alone can not provide enough
attraction to stop the inflation.
Obviously one need an extra attractive force to end the exponential
expansion 
(as well as an extra repulsive force to drive the inflation),
 as shown in Fig.1.
Again we may have to generalize the Einstein's theory,
if we are to accommodate this extra interaction.

All these arguments, although mutually independent, suggest the 
existence of a fundamental scalar field which we call the dilaton
which could affect the gravitation (and consequently the cosmology)
in a fundamental way.
In the following we  discuss how the dilaton comes about in the 
unified field theory, and how it could affect the gravitation and
cosmology.

\vskip 1cm
\noindent {\bf II. Gravitation and Unified Field Theory}
\vskip 0.7cm

To see how the dilaton appears in the unified theory, let us
first consider the $(4+n)$--dimensional Kaluza-Klein
theory whose fundamental ingredient is
the $(4+n)$--dimensional metric $g_{AB}$ $(A,B=1,2,\cdots,4+n)$
\begin{eqnarray}
g_{AB}=\left(
\begin{array}{cc}
 \tilde{g}_{\mu\nu}+e_0\kappa_0 \phi_{ab}A^a_\mu A^b_\nu &
 \hspace{3mm} e_0 \kappa_0 A_\mu^a \phi_{ab} \\
 e_0\kappa_0 \phi_{ab}A^b_\nu & 
 \hspace{3mm} \phi_{ab}
\end{array}
\right),
\end{eqnarray}
where $e_0$ is a coupling constant, and $\kappa_0$ is a scale parameter
which sets the scale of the $n$-dimensional internal space.
When the metric has an $n$-dimensional  isometry $G$, one can reduce the 
$(4+n)$-dimensional Einstein's theory to a $4$-dimensional unified
theory\cite{cho,cho4}. 
Indeed with $e_0^2 \kappa_0^2 = 16 \pi G$ and with
\begin{eqnarray}
&&\tilde{g} = |{\rm det} ~\tilde{g}_{\mu \nu}| , 
\hspace{8mm} \phi =|{\rm det}~\phi_{ab}|,\nonumber \\
&&\rho_{ab} = \phi^{-\frac{1}{n}}\phi_{ab},
\hspace{8mm}(|{\rm det}~\rho_{ab}|=1),
\end{eqnarray}
the $(4+n)$-dimensional Einstein's theory is reduced to the following 
$4$-dimensional Einstein-Yang-Mills theory,
\begin{eqnarray}
\label{lag1}
{\cal L}_0 = -\frac{1}{16\pi G}\sqrt{\tilde{g}}
\sqrt{\phi}~\Bigl[&&\tilde{R} + \tilde{S} \nonumber \\
&&-\frac{n-1}{4n}\frac{(\partial_{\mu}\phi)^2}{\phi^2}
+\frac{1}{4}\rho^{ab} \rho^{cd}(D_{\mu}\rho_{ac})(D_{\mu}\rho_{bd})
+\Lambda+ \lambda(|{\rm det}~\rho_{ab}|-1)\nonumber \\
&&+ 4\pi G\phi^{\frac{1}{n}}\rho_{ab}F_{\mu\nu}{}^aF_{\mu\nu}{}^b+\cdots
   \Bigr],
\end{eqnarray}
where $\tilde{R}$ and $\tilde{S}$ are the scalar curvature of
 $\tilde{g}_{\mu\nu}$ and $\phi_{ab}$, $\Lambda$ is the $(4+n)$-dimensional
 cosmological constant, $\lambda$ is a Lagrange multiplier.
But notice that the above Lagrangian has a crucial defect. First the metric 
$\tilde{g}_{\mu\nu}$ does not represent 
the Einstein metric because $\tilde{g}$
does not describe the proper $4$-dimensional volume element. 
But more seriously the $\phi$-field appears with a negative 
kinetic energy, and thus can not be treated as a physical field \cite{cho}.
To cure this defect one must perform the following conformal transformation,
and introduce the physical metric $g_{\mu\nu}$ and the dilaton field
$\sigma$ by
\begin{eqnarray}
&&g_{\mu\nu} = \sqrt{\phi}~\tilde{g}_{\mu\nu},\nonumber \\
&&\sigma = \frac12 \sqrt{\frac{n+2}{n}}\ln \phi.
\end{eqnarray}
With this the Lagrangian (\ref{lag1}) is written as 
\begin{eqnarray}
\label{lag2}
{\cal L}_0 = 
- \frac{\sqrt{g}}{16\pi G}\Bigl[&& R 
+ \frac12 (\partial_{\mu}\sigma)^2 \nonumber  \\ 
& & + S e^{-\alpha\sigma} + \Lambda e^{-\alpha^{'}\sigma}
-\frac14 (D_{\mu}\rho^{ab})(D_{\mu}\rho_{ab}) 
+\lambda(|{\rm det}~\rho_{ab}|-1) \nonumber \\
&&+ 4\pi G e^{\alpha\sigma}\rho_{ab} F_{\mu\nu}{}^aF_{\mu\nu}{}^b + \cdots 
\Bigr],
\end{eqnarray}
where $R$ and $S$ are the scalar curvature of $g_{\mu\nu}$ and $\rho_{ab}$,
$\alpha$ and $\alpha^{'}$ are the coupling constants given by 
$\alpha = \sqrt{(n+2)/n}$ and $\alpha^{'} = \sqrt{n/(n+2)}$ .
This shows that in the Kaluza-Klein theory the dilaton appears 
as the volume element of the internal metric
which, as a component of the metric $g_{AB}$, must couple to all
matter fields.

 In superstring theory the dilaton appears 
as the massless scalar field that
the mass spectrum of the closed string must contain.   
After the full string loop expansion, the $4$-dimensional
effective Lagrangian of the massless modes has the following form 
in the string frame\cite{fra,Da} 
\begin{eqnarray}
\label{stlag1}
{\cal L}_S &=& 
-\frac{\sqrt{\tilde{g}}}{\kappa}
\Bigl[ \tilde C_g(\varphi)\tilde{R}
+\tilde C_{\varphi}(\varphi)(\partial_{\mu}\varphi)^2 
-\frac{\kappa}{4}\tilde C_1(\varphi) (F_{\mu\nu}{}^a )^2
+\cdots \Bigr],
\end{eqnarray}
where $\kappa$ is the string slope parameter, 
$\tilde g_{\mu\nu}$ is the string frame metric, 
$\varphi$ is the string dilaton,
 and $\tilde C_i(\varphi)~(i= g,\varphi,1,2,3,\cdots)$ are
 the dilaton coupling
functions to various fields.
 At present their exact forms are not known 
beyond the fact that in the limit $\varphi$ goes to $-\infty$
they should admit the following loop expansion
\begin{eqnarray}
\label{scf}
\tilde C_i(\varphi)=e^{-2\varphi}+a_{i} + b_{i} e^{2\varphi}
+c_{i} e^{4\varphi} + \cdots .
\end{eqnarray}
Notice that the effective Lagrangian~(\ref{stlag1}) looks
 quite similar to the Lagrangian~(\ref{lag1}) which
 we obtained from
the Kaluza-Klein theory. Indeed, introducing
a new metric $g_{\mu\nu}$ with the conformal 
transformation
and replacing the original dilaton field $\varphi$ with a new  
one $\sigma$ by
\begin{eqnarray}
& & g_{\mu\nu}=\tilde C_g(\varphi) \tilde g_{\mu\nu}, \nonumber \\
& & \sigma =\int \Bigl[
~\frac{3}{4}\frac{1}{\tilde C_g}\frac{d \tilde C_g}{d \varphi} 
+2\frac{1}{\tilde C_g}\frac{d \tilde C_{\varphi}}{d\varphi}
+2\frac{\tilde C_{\varphi}}{\tilde C_g}
~\Bigr]^{1/2}d\varphi,
\end{eqnarray}
one may put the Lagrangian~(\ref{stlag1}) into the
 following standard form
\begin{eqnarray}
\label{stlag2}
{\cal L}_S =
\frac{\sqrt{g}}{\kappa}\Bigl[
R+\frac{1}{2}(\partial_\mu \sigma)^2
+\frac{\kappa}{4}C_1(\sigma) F_{\mu\nu}{}^a F_{\mu\nu}{}^a 
+\cdots \Bigr].
\end{eqnarray}
Notice that in the standard form
the dilaton coupling function to gravity  $\tilde C_g(\varphi)$ and
 the self coupling function $\tilde C_{\varphi}(\varphi)$ 
disappear completely with the 
redefinition of the fields. Only the coupling functions
to the other matter fields remain.

Now the Lagrangian (\ref{stlag2}) looks almost
 identical to the Lagrangian (\ref{lag2}) of
 the Kaluza-Klein theory. In both
 cases the dilaton appears as a fundamental scalar field.
Of course there are some differences.
One of them is the form
 of the dilatonic coupling functions to various matter
fields. In the Kaluza-Klein
 theory they have simple exponential
forms, whereas in the string theory
 their explicit forms are not known. 
Another is the mass of the dilaton.
In the Kaluza-Klein  theory the dilaton can easily acquire a mass,
but in the superstring theory it remains masslss to all orders of
perturbation\cite{fra,Da}.
But these differences is may not be so serious as it appears. 
To understand this one has to keep in mind
two things. First, the Lagrangian~(\ref{lag2}) in
 the Kaluza-Klein theory
is valid only at the tree level, because it
 did not take into account the full
renormalization effect. 
 Moreover (\ref{scf}) shows
 that, at the tree level
in the string loop expansion, the string coupling
 functions also have the
exponential forms. So with the quantum correction
 in the Kaluza-Klein
theory, the difference in the dilaton coupling functions
 between the two
theories  could become insignificant. 
As for the mass of the dilaton, there is no fundamental principle
which can keep it massless, even enough the perturbative expansion 
leaves it massless in the string theory. So it could acquire a mass though
some unknown non-pertubative or topological mechanism.
From these one may conclude that {\em as far as the
dilaton is concerned the string theory
 and the Kaluza-Klein theory
give us practically the same effective Lagrangian, at
 least in the
low energy approximation}.
In both cases the dilaton comes into
 play an important ingredient as
 the spin-zero partner of the spin-two
 graviton which is responsible for the
Einstein's gravitation. So in the unified
 theory one must take the dilatonic
modification of the Einstein's theory
 seriously, whether one likes it or not. 

\vskip1.0cm
\noindent{\bf III. Brans-Dicke Theory}
\vskip0.7cm

One may wonder why the string theory and the
 Kaluza-Klein theory give us
almost identical effective Lagrangian. There
 is a good reason for
this. To understand this it is instructive to
 discuss the Brans-Dicke
theory first. In an attempt to generalize the Einstein's
 gravitation with a fundamental
scalar field, Brans and Dicke arrived at the
 following Lagrangian \cite{brans}
\begin{eqnarray}
\label{bdlag1}
{\cal L}_{BD} &=& {\cal L}_{g} + {\cal L}_{m}, \\
\cr
{\cal L}_{g} &=& -\sqrt{\tilde g} \hspace{2mm} \Bigl[
\phi \tilde R +\frac{\omega}{\phi}(\partial_\mu \phi)^2 \Bigr], \\
\cr
{\cal L}_{m} &=& 
-\sqrt{\tilde g} \hspace{2mm} \Bigl[ 
\frac{1}{4} (F_{\mu\nu})^2
+ \overline{\psi}i \gamma^{\mu} D_{\mu} \psi
- m \overline{\psi}\psi \Bigr],  
\end{eqnarray}
where $\tilde g_{\mu\nu}$ is the Jordan metric, $\phi$ is
 the Brans-Dicke scalar field,
and $\omega$ is the coupling constant. Notice that here
 we have kept the electromagnetic field $F_{\mu\nu}$ and 
a fermion field $\psi$ as the matter fields for simplicity. But
 the important point here is that in the Jordan frame the dilaton
 coupling functions to the matter fields (other than the graviton and 
the dilaton) are all chosen to be trivial, so that the dilaton does not
 couple directly to the matter fields. 
 In other words the ordinary matter is allowed to couple to
 the gravitation only ``minimally" through the Jordan metric, 
 in spite of the fact that the Brans-Dicke scalar field
 is an important element of gravitation. What is remarkable is that, 
if we change the Jordan frame to another with a conformal
 transformation, this minimal coupling no longer holds. To
 see this let us introduce
 the Pauli metric $g_{\mu\nu}$ and the Brans-Dicke dilaton $\sigma$ 
 by\cite{cho1}
\begin{eqnarray}
\label{fourteen}
&&g_{\mu\nu}= e^{\alpha\sigma}\tilde g_{\mu\nu}\ , \nonumber \\
&&\phi=\frac{1}{16\pi G} e^{\alpha\sigma},
\end{eqnarray}
where
\begin{eqnarray}
\label{alpha}
\alpha = \frac{1}{\sqrt{2\omega +3}}.
\end{eqnarray} 
Now in the Pauli frame one can easily show that 
 the Brans-Dicke Lagrangian acquires the following standard form,
\begin{eqnarray}
\label{bdlag2}
{\cal L}_{BD} =&&- \frac{\sqrt{g}}{16\pi G} \hspace{2mm} \Bigl[ 
R+\frac{1}{2}(\partial_\mu \sigma)^2 \Bigr] \nonumber \\  
&&- \sqrt{g} \hspace{2mm} 
\Bigl[ \frac{1}{4}(F_{\mu\nu})^2 + 
e^{-{2 \over 3} \alpha\sigma } \bar{\psi} i \gamma^{\mu} D_{\mu}\psi -
e^{-2\alpha\sigma}m\bar{\psi}\psi 
\Bigr].
\end{eqnarray}
So the dilaton 
coupling function acquires  a non-trivial form
  in the Pauli frame (except for
 the $U(1)$ gauge field), so that now the dilaton has
 a direct coupling to the matter field. The reason why the coupling 
function to the gauge field remains trivial is simply 
because the gravitational interaction of the gauge field 
is conformally invariant under (\ref{fourteen}).

Brans and Dicke has argued that the above theory 
 is the only acceptable theory of gravitation 
containing a massless 
scalar graviton which respects the weak equivalence principle. 
But it is clear that Lagrangian (\ref{bdlag1})
 is a special case of (\ref{stlag1}),
 in which all the dilaton coupling
 functions are uniquely determined. 
The reason why they have chosen
 the above coupling functions was very simple.
 The equivalence principle requires a test
 particle made of the matter fields
 to follow geodesics determined by the physical metric. 
To guarantee this the physical metric must couple {\em minimally} 
to the matter fields, which means that the dilaton
 should not couple directly to the matter field.
 This requires the dilaton
coupling functions to the matter fields to be trivial.
 Of course Brans and Dicke identified
 the Jordan metric as physical, 
and thus arrived at the Lagrangian (\ref{bdlag1}).

This tells us the followings. First, there is practically 
 one way to generalize the     Einstein's theory 
with a (massless) scalar graviton. 
The only freedom one can have is the
 dilaton coupling functions to the
matter fields. This is why all the above
 theories---the Brans-Dicke theory,
 the Kaluza-Klein theory, and the superstring theory---give
us practically the same effective 
 Lagrangian.
The other point is that the form of
 dilaton coupling functions
depends on the conformal frame that one chooses.
For example, in the Brans-Dicke theory
 the dilaton coupling functions 
are trivial in the Jordan frame, but acquire a non-trivial
form in the Pauli frame. This means that
  a test particle will follow
 a geodesic in the Jordan frame, but not in the Pauli frame.
So it is crucial for us to decide what is the physical frame 
before we discuss the physics. 
Considering the fact that  the conformal 
 invariance is clearly broken in the real world, 
this is perhaps what one 
should have expected. Nevertheless we 
find that this point is not so well 
appreciated in the literature.

As we have emphasized, the equivalence principle
 was the underlying principle for the Brans-Dicke theory.
 But can one really maintain the equivalence
 principle in the Bran-Dicke theory? 
 Brans and Dicke argued that one can do so,
 if (and only if) one treats the Jordan metric as physical.
 To guarantee this they have forbidden
 a direct coupling of Brans-Dicke scalar field
 to the ordinary matter in the Jordan frame,
 even though the scalar field was an essential
 ingredient of gravitation.
 Unfortunately this does not guarantee
 the equivalence principle in the Brans-Dicke theory \cite{apctp}.
 This is because the minimal coupling of
 the Jordan metric to the ordinary matter becomes 
unstable under the quantum fluctuation.
Indeed, when the quantum correction takes place,
 the ordinary matter must couple to
 the Brans-Dicke scalar field through 
 the Jordan metric, as shown in Fig.2. 
So the quantum fluctuation inevitably induces
 a direct coupling of the Brans-Dicke scalar
 field to the ordinary matter, even in the Jordan frame. 
This tells that, when the quantum correction  takes place, 
there is no way to
 enforce the equivalence principle 
in the Brans-Dicke theory.

The Lagrangian (\ref{bdlag2}) tells that
 in the Pauli frame the dilaton couples
 directly to the charged scalar field,
 but not to the electromagnetic field. Naively this
 would imply that the charged particle
 will not follow the geodesic, 
but the photon will.
 This appearance, however, is misleading because
 the quantum fluctuation must necessarily
 induce a direct coupling of the dilaton to the gauge field. 
Indeed the quantum correction shown
 in Fig.3 should add an induced interaction \cite{apctp} 
\begin{eqnarray}
\label{quamflu}
\delta {\cal L}\simeq  \alpha \alpha_{e} \sigma F_{\mu\nu}F_{\mu\nu}, 
\end{eqnarray}
to the Lagrangian (\ref{bdlag2}). 
Clearly this induced
 coupling is suppressed by
 the factor $\alpha_e$, compared to 
the direct coupling which
 already exists at the classical level. 
 This shows that the dilaton couples to different matters with 
different strengths. It is this
 ``composition-dependent" coupling that
 violates the equivalence principle
 in the Brans-Dicke theory.
       
 Since the Brans-Dicke theory is a prototype 
theory of gravitation that one finds in
 all existing unified theories, it is 
important to understand the theory
 in more detail. We emphasize a few 
characteristics of the Brans-Dicke theory :\\
1) It is the Pauli metric, not
 the Jordan metric, which describes the massless 
spin-two graviton and thus
 the Einstein's gravitation\cite{cho1}.
 In fact the Jordan 
metric is a strange mixture of the spin-two graviton
 and spin-zero dilaton which
does not even describe a mass
 eigenstate. This tells that,
 when one wants to 
compare the theory with the Einstein's gravitation,
 one must use the Pauli frame.
Furthermore, when one tries to quantize the theory,
 obviously the Pauli frame
 comes in as the natural frame.
 So from the logical point of view the 
Pauli frame becomes the most natural
 frame to discuss the physics.

2) In the Pauli frame the Brans-Dicke dilaton
 describes a scalar component of gravitation
 which is absent in the Einstein's gravitation. Naturally
 this ``new" gravitation could be interpreted as the dilatonic 
``fifth force" which modifies the Einstein's gravitation.
 In this view the huge Brans-Dicke coupling constant 
$\omega\ (\omega >600)$ 
translates to a perfectly reasonable
 new constant $\alpha$ $(\alpha <0.03)$
 through (\ref{alpha}), which determines the coupling
 strength of the dilatonic fifth force to the ordinary
 matter. This tells that the Brans-Dicke
 theory is really a theory of fifth force.

The importance of the above discussion is
 that these characteristic features of Brans-Dicke
 theory, in particular the existence of the dilatonic
fifth force and the violation of the
 equivalence principle, 
 should also apply to the Kaluza-Klein theory
 and the superstring theory. The only
 difference is that the situation gets worse in the
unified theories. 
For example, in these theories
 the violation of equivalence principle
 takes place already at the classical
 level. This is because here the dilaton
 couples to different matter with
different strengths even without  any quantum correction.  
 To make the matter worse, in the string theory the dilatonic 
fifth force becomes intolerably large, 
because here the massless dilaton couples as strongly as the gravitation 
 (with $\alpha \simeq 1$) to the matter field.  

\vskip1.0cm
\noindent{\bf IV. Dilatonic Fifth Force}
\vskip0.7cm

Now we discuss the dilatonic fifth force \cite{cho1,cho2} in a general
setting. We have shown that the dilatonic
coupling constant $\alpha$ may in principle
 depend on the type of
matter field it couples, when there are more
 than one type of matter fields
in the theory. 
But for simplicity one may assume that only one type of
coupling, the dilatonic coupling
 to the baryonic matter, is important
for the practical purpose. Given the fact that the
 baryonic matter is the only dominant
 matter of the universe verified so far, the assumption
 is well justified.  In this case
 only one universal coupling
constant $\alpha$ characterizes the fifth force. With this
the important issue now becomes
how strong is the fifth force, and how far does it act. The
strength is determined by
the coupling constant $\alpha$, but the range
is determined by the mass $\mu$ of
the dilaton. Let $F_g$ and $F_5$ be the
 gravitational and the fifth force
between the two mass points $m_1$
and $m_2$ separated by $r$. From the
 dimensional argument one may
express the total force $F$
in the Newtonian approximation as
\begin{equation}
F = F_g + F_5 \simeq \frac{\alpha_g}{r^2}+\frac{\alpha_5}{r^2}e^{-\mu r}
  = \frac{\alpha_g}{r^2} (1+\beta e^{-\mu r}) ,
\label{7}
\end{equation}
where $\alpha_g$ and $\alpha_5$
are the fine structure constants of
 the gravitation and the fifth
force, and $\beta$ is the ratio
 between $\alpha_g$ and $\alpha_5$.
In terms of Feynman diagram the first term
 represents one graviton exchange but the second
 term represents one dilaton exchange in the zero
momentum transfer limit.
Notice that in the Brans-Dicke theory the Lagrangian (\ref{bdlag2}) suggests
\begin{equation}
\beta \simeq \alpha^2,
\label{8}
\end{equation}
from which one can easily estimate $\beta$. 
For instance $\omega > 600$ with (\ref{alpha}) implies 
$\beta \simeq \alpha^2 < 10^{-3}$.

To proceed further one must know the mass of the dilaton.
Of course the dilaton appears massless in the superstring and
the Brans-Dicke theory.
If the dilaton remains strictly massless,
 there is no way to differentiate the fifth
 force from the gravitation in this Newtonian approximation.
 This is because the net effect of the massless 
 dilaton is simply to replace $\alpha_{g}$ with
 the effective gravitational
 constant $\overline{\alpha}_g=(1+\beta) \alpha_{g}$.
In the absence of any simple mechanism which can keep
 the dilaton massless, however,
it is reasonable to assume that the dilaton acquires 
a small mass through some unknown
 quantum correction. Unfortunately
 it is extremely difficult to estimate
this quantum correction at present.
Under this circumstance one
 may leave $\mu$ as a free parameter and consider the following
possibilities :\\
a) $\mu \simeq 0$ (long range).
From the existing experimental data on the long range 
fifth force we have $\beta < 10^{-9}$ \cite{fish}.
Notice that in the Brans-Dicke
 theory $\beta < 10^{-9}$ amounts to $\omega > 10^{8}$, which gives us
a much more stringent constraint on the Brans-Dicke coupling
constant than the existing
 bound $\omega > 600$\cite{will}. 
 Clearly the new bound is made possible
by the interpretation that the Brans-Dicke theory is really a theory
of a fifth force in which the new gravitational force is generated by
the Brans-Dicke scalar field. Of course such a large $\omega$ (or
such a small $\beta$) might be interpreted to imply that there
is no such long range fifth force. At this point, however, it may
be good to remember that the ratio between the gravitational and
the electromagnetic coupling of the elementary particles is
extremely small, $\alpha_g/\alpha_e\simeq10^{-40}$.\\
b) $\mu \simeq 2\times10^{-10}$ eV (1 km range).
In this medium range we have $\beta e^{-\mu r}\leq 10^{-4}$ 
experimentally. 
This is perfectly consistent with our 
estimate of $\beta$ based on $\omega > 600$ in the 
Brans- Dicke theory. 
Of course $\beta$
could still be much smaller than $10^{-4}$ here, in which
case a best way to measure
$\beta$ is the laboratory (small size) experiments.\\
c) $\mu\simeq1$ keV ($2\times10^{-8}$ cm range).
In this atomic scale there is no experimental
 constraint yet, and the possibility of
$\beta\simeq1$ can not be ruled out.
 In fact all the unified theories predict $\beta$ to be of the order one.
This case is particularly interesting because a 0.5 keV dilaton
could be an excellent candidate of the dark matter 
of the universe, as we will discuss in the following. \\
d) $\mu\simeq10^{15}$ GeV ($2\times10^{-28}$ cm range).
In this grand unification scale there is
 practically no way to detect the fifth force in the
present universe, even though it
 may very well exist at this short distance.
Notice, however, that in the early universe
 this fifth force could have played an
important role to stop the inflation by
 providing an extra attractive
force to curb the exponential 
expansion of the universe \cite{apctp}.

The above analysis teaches us the followings. 
First, 
the modification of the Einstein's
 gravitation by the massless dilaton (a long range fifth force)
 has to be extremely small, if there is any.
Indeed the weak dilatonic coupling to the ordinary matter field restricts
 $ \omega > 10^8 $ in the Brans-Dicke theory, 
which sets the most stringent bound
 for the Brans-Dicke coupling constant. 
Furthermore this extreme weak dilatonic coupling must apply 
to all unified theories, as far as the dilaton remains
massless. So in any theory with a massless dilaton one
must find a theoretical justification why the dilatonic coupling
 is so weak. This (together with $\alpha\simeq1$)
creates a serious problem for the superstring
theory, where the string dilaton remains massless to all orders
of perturbation\cite{Da,taylor}. Secondly, a fifth force by
a massive dilaton, even with a relatively small mass of $10^{-10}$ eV
range, is very difficult to rule out experimentally.
In fact there is practically no hope to detect 
a dilatonic fifth force at about the grand unification scale
 in the present universe. 
Nevertheless this provides a most interesting
possibility from the cosmological point of view, because 
this type of fifth force may have played a crucial role 
in the evolution of
the universe at the early stage. 
Furthermore the massive dilaton could provide an excellent candidate
for a non-baryonic dark matter.
\vskip1.0cm
\noindent{\bf V. Unified Cosmology}
\vskip0.7cm

It is generally believed that, to solve 
the major problems of the standard cosmology, one
may need an inflation at the early stage of universe.
A best way to implement the inflation is to introduce a
scalar field as the inflaton field.
 But this inevitably leads us to a ``generalized" Brans-Dicke
theory, which is exactly what we find in the unified field theories
 as we have discussed in the above. This implies that the unified field
theories can naturally provide us an inflation\cite{cho5}. 
In fact the possibility of an inflation in all the unified
 theories has been successfully
argued by many authors\cite{Da,dla}.
This is because in these theories
the dilaton could 
 assume the role of the inflaton, so that by choosing a proper
inflationary potential one could obtain a successful inflation.

There is, however, an important point that one must keep
in mind when one tries to implement
 an inflation in these theories.
To discuss the inflation one must first
 decide which conformal frame is
the physical frame. This is because the
 actual expansion rate of
inflation depends on the conformal frame that one chooses\cite{cho5}.
Unfortunately many of the authors in the literature have
overlooked this important point, and
 have used unphysical frames without
 a proper justification.

Depending on the dilatonic potential the unified theory
provides us with a wide range of cosmology. But there are a few
characteristics of the unified cosmology\cite{cho5}: \\
a) The matter in the unified cosmology consists of two parts,
the dilatonic matter and the ordinary matter.
Futhermore the density $\rho_{\sigma}$ of the dilatonic matter 
is generically given by
\begin{equation}
\rho_{\sigma} \simeq {1 \over 16 \pi G} {H^2 \over \alpha^2}
\simeq \rho_c
\end{equation}
so that $\Omega_{\sigma}$ could become of the order one.
This suggests that the dilaton could easily become 
the dark matter of the universe.

b) In the unified cosmology the Dirac's conjecture is realized,
but the time-dependence of the Newton's constant is given by
\begin{equation}
{\dot G \over G} \simeq H \simeq 10^{-10}/year.
\end{equation}
So it could naturally accomodate the present
experimental constraint on $\dot G/G$.

\vskip1.0cm
\noindent{\bf VI. Dilatonic Dark matter}
\vskip0.7cm

The unified cosmology implies that the dilaton could
 be the dark matter of the universe. In this section
we first show that the dilaton starts with the thermal
equilibrium from the beginning and decouples from the
 other sources very early near the Planck scale.
 After the decoupling the fate of the dilaton crucially
depends on the mass. We find that there are two mass ranges,
$\mu \simeq$ 0.5 keV and $\mu \simeq$ 270 MeV, 
where the dilaton could be the dominant matter of the universe.
The dilaton with mass larger than 270 MeV 
does not survive long enough to become the dominant
matter of the present universe, and the
dilaton with mass smaller than 0.5 keV  survives but
fails to be dominant due to the low mass.
 The dilaton with mass in between can not be seriously
 considered because it would overclose the universe. Remarkably
 the 0.5 keV dilaton has the free-streaming
distance of about 1.4 Mpc and provide an excellent
candidate of a warm dark matter, but the 270 MeV dilaton
has much shorter free-streaming distance of about 7.4 pc
so that it becomes a cold dark matter.

To show how the dilaton reaches the
thermal equilibrium from the beginning notice that the dominant
interaction modes of the dilaton with other matter fields are the
Feynman diagrams shown in Fig.4. Normally the dilatonic coupling
strength would be $\alpha\, m_q/m_p$, where $\alpha$ is the 
dimensionless coupling constant, 
$m_p$ is the Planck mass, and $m_q$ is the mass of
the dominant matter (the quarks). 
But notice that at high temperature
(at $T\gg m_q$), the coupling strength becomes $\alpha\, T/m_p $.
 With this one can easily estimate the dilaton creation (and
annihilation) cross section shown in Fig.4 (a)
\begin{equation}
\sigma \simeq g^2\alpha^2 \Bigl(\frac{T}{m_p} \Bigr)^2
\times\frac{1}{T^2},
\end{equation}
so that the creation rate $\Gamma$ is given by
\begin{equation}
\Gamma \simeq n_q\sigma v \simeq g^2 
\alpha^2\Bigl(\frac{T}{m_p}\Bigr)^2\times T.
\end{equation}
Similarly the scattering cross section $\sigma$ shown in Fig.4 (b)
 is given by
\begin{equation}
\sigma \simeq \alpha^4 \Bigl(\frac{T}{m_p}\Bigr)^4 \times\frac{1}{T^2}
\end{equation}
with the following interaction rate $\Gamma$
\begin{equation}
\Gamma \simeq n_q\sigma v \simeq \alpha^4 
\Bigl(\frac{T}{m_p}\Bigr)^{4} \times T.
\end{equation}
On the other hand the Hubble expansion rate $H$ in
the early universe is given by\cite{turner}

\begin{equation}
H \simeq \frac{T^2}{m_p}.
\end{equation}
From this we conclude that {\it the dilaton is thermally
produced from the beginning, and decouples with the other sources
at around the Planck scale with the decoupling temperature $T_d$
given by}
\begin{equation}
T_d \simeq \frac{m_p}{\alpha^{4/3}}.
\end{equation}
Notice that the dilaton decouples with the other sources at around the
same time as the graviton does. This is indeed what one would
have expected, since the dilaton is nothing but the scalar
 counterpart of the Einstein's graviton.

Once the dilaton acquires a mass, it becomes unstable and
 decays to the ordinary matter. A typical
decay process is the two photon process and the fermion pair production
process described by the following interaction Lagrangian
\begin{eqnarray}
{\cal L}_{int}\simeq &&- \frac{\alpha_1}{4} \sqrt{16\pi G}\, \phi
F_{\mu\nu}F_{\mu\nu}\nonumber \\
 &&-\alpha_2  \sqrt{16\pi G}m\,\phi\,\bar{\psi}\,\psi.
\end{eqnarray}
For the two photon process we obtain the following life-time at
 the tree level\cite{cho-keum}
\begin{equation}
\tau_1 \simeq \frac{16}{\alpha_1^2} \Bigl(\frac{m_p}{\mu}\Bigr)^2
\frac{1}{\mu} \hspace{3mm} \Bigl(\times\frac{T}{\mu}\Bigr),
\end{equation}
where the last term in parenthesis is the time-dilatation effect which becomes
important only at high temperature when the dilaton
becomes relativistic. Similarly for the pair production we obtain

\begin{eqnarray}
\tau_2 \simeq && \frac{1}{2 \alpha_2^2}
\Bigl(1-4\frac{m^2}{\mu^2}\Bigr)^{-3/2}\,
\Bigl(\frac{m_p}{m}\Bigr)^2 \, \frac{1}{\mu} \, 
\hspace{3mm} 
\Bigl(\times \frac{T}{\mu}\Bigr)
\nonumber \\
  \geq && \frac{5.38}{\alpha_2^2} \Bigl(\frac{m_p}{\mu}\Bigr)^2 \,
\frac{1}{\mu
}\, \hspace{3mm} \Bigl(\times\frac{T}{\mu}\Bigr).
\end{eqnarray}
A more detailed calculation which includes all possible decay
 channels allowed in the standard electroweak model 
 gives us Fig.5
of the dilaton life-time with respect to the dilaton mass\cite{cho-keum}.
Notice that here we have assumed
$\alpha_1 \simeq \alpha_2 \simeq 1$ for simplicity, 
but one should keep in mind that in reality the coupling constants
could turn out to be much smaller.

To estimate how much the dilaton contributes to
 the matter density of the present universe
one must estimate the number density of the
dilaton at present time. From the entropy
conservation of the universe one
can easily estimate the present temperature $T_{\phi}$ of
the dilaton. 
Based on the standard electroweak theory
one finds
\begin{equation}
T_{\phi}\leq \Bigl(\frac{3.91}{106.75}\Bigr)^{1/3}\, 
T_0 \simeq 0.91^{\circ} K,
\end{equation}
where $T_0$ is the present temperature of the background radiation.
Notice that again this is the temperature of the graviton
at present time. From this one can estimate the number density $n_0$
 of the dilaton at present. Assuming that
 the dilaton is stable one has
\begin{equation}
n_0 = \frac{\zeta (3)}{\pi^2}\,T_{\phi}^3 \simeq 7.5\, /cm^3.
\end{equation}
But as we have emphasized, the massive dilaton can not
be stable, and the number density
of the dilaton $n(\mu)$ must crucially depend on
its mass. So for the dilaton to provide the
critical mass of the universe one must have
\begin{equation}
\rho (\mu) = n(\mu) \times \mu = n_0 \, 
e^{-t_0/\tau(\mu)}\times \mu \simeq 
10.5 \hspace{2mm}h^2 \hspace{2mm} keV/cm^3.
\end{equation}
where $t_0$ is the age of the universe,
 $\tau(\mu)$ is the life-time of the dilaton,
and $h$ is the Hubble constant (in the unit of 100Km/sec Mpc).
A numerical calculation with $t_0 \simeq 1.5 \times 10^{10} $ years 
 shows that 
{\it there are two mass ranges, 
$\mu \simeq 0.5$ keV 
or $\mu \simeq 270$ MeV,
which can make the dilaton a candidate of the dark matter in the universe.}
In Table I the interesting physical quantites are shown for
 different values of $h$.

Notice that with $h \simeq 0.6$ the mass becomes
0.5 keV or 270 MeV.
Also notice that the $\rho(\mu)$ starts from zero when $\mu = 0$ and
reaches the maximum value at 0.5 keV $< \mu < $ 270 MeV and
again decreases to zero when $\mu = \infty$. This means
that when $\mu < $ 0.5 keV or $\mu > $ 270 MeV
 the dilaton undercloses
the universe, but when 0.5 keV $ < \mu < $ 270 MeV 
it overcloses the universe. 
From this one may conclude that 
{\em the dilaton with 0.5 keV $ < \mu < $ 270 MeV 
is not acceptable because this is incompatible with the cosmology.} 
In view of the fact that the dilaton must exist in all
 the unified field theories, the above constraint on
the mass of the dilaton should provide us an important piece of
information in search of the dilaton.

Now we discuss the possibility
of the dilatonic dark matter in more detail : \\
a) $\mu \simeq $ 0.5 keV. 
In this case the available decay channel is the
  $ \gamma\gamma $ process.
So the life-time is given by $\tau \simeq 4.0 \times 10^{26}$ years,
which tells that it is almost stable. To determine whether this
dilaton could serve as a hot or cold dark matter, one
must estimate the free-streaming distance $\lambda$ of
the dilaton. The dilaton becomes non-relativistic around
 T$\simeq \mu/3\simeq 0.17$ keV,  
long before the matter-radiation equilibrium era.
In terms of time this corresponds to \cite{turner}
\begin{equation}
t_{NR}\simeq 1.2 \times 10^7\times
\Bigl(\frac{keV}{\mu}\Bigr)^2\,
\Bigl(\frac{T_\phi}{T_0}\Bigr)^2\, sec \simeq 
1.88 \times 10^{6} \, sec.
\end{equation}
From this one obtains
\begin{eqnarray}
\lambda \simeq&& 0.16\,
\Bigl(\frac{keV}{\mu}\Bigr)\Bigl(\frac{T_\phi}{T_0}\Bigr
) \bigg[ \ln\Bigl(\frac{t_{EQ}}{t_{NR}}\Bigr) +2\bigg]\, Mpc\nonumber\\
\simeq&& 1.4 \, Mpc.
\end{eqnarray}
Certainly this is a very interesting number, 
which tells that the 0.5 keV
dilaton becomes an excellent candidate of a warm dark matter.\\
b)
$\mu\simeq 270$ MeV. 
In this case the available decay channels
are the $\gamma\gamma$,
$e^{+} e^{-}$, and $\mu^{+} \mu^{-}$ processes
(the $\nu\bar{\nu}$ processes are assumed to be negligible).
The decay processes $\gamma\gamma, \mu^{+}\mu^{-}$ are
dominant at this energy level, and have almostly the same decay width (See Table I).
The corresponding life-time is
given by $\tau\simeq 1.1 \times 10^9$ years,
 so that only a fraction of the thermal dilaton survives now. 
For this dilaton one has
\begin{equation}
t_{NR}\simeq 1.82 \times 10^{-5} \, sec,
\end{equation}
and the corresponding free-streaming distance becomes
\begin{equation}
\lambda \simeq 7.35 \, pc.
\end{equation}
Clearly  this dilaton becomes a good
candidate for a cold dark matter.

Now the important question is how one could detect the dilaton. It
seems very difficult to detect it through the dilatonic fifth force,
because the range of the fifth force would be about
$10^{-8}$ cm (for $\mu = 0.5$ keV) or 
about $10^{-13}$ cm (for $\mu = 270$ MeV).
Perhaps a more promising way is to use the two photon decay process,
which produces two mono-energetic X-rays of
 $ E \simeq 0.25$ keV or 
$E \simeq 135$ MeV with the same polarization. 
With the local halo density of our galaxy 
$\rho_{HALO}\simeq 0.3$ GeV/$cm^3$
 one can easily find the local dilaton number density to be
$\bar n\simeq 5.83 \times 10^5 /cm^3$ for $\mu = 0.5$ keV 
and $\bar n\simeq 0.11 /cm^3$ for  $\mu = 270$ MeV. 
In both cases the local velocity of the dilaton is about $10^{-3}$ c.
So it is very important to look for the above X-ray signals
from the sky
(with the Doppler broadening of $\Delta E\simeq 10^{-3}E$)
or to perform a Sikivie-type X-ray detection experiment
with a strong electromagnetic field to enhance the dilaton conversion,
although the long life-time (for $\mu = 0.5$ keV) or 
the low local number density (for $\mu = 270$ MeV)
of the dilaton could make
such experiments very difficult.
For the  $\mu$ = 270 MeV dilaton  
one could also look for the $\mu^{+}\mu^{-}$ decay process.

One might try to detect the dilaton from the accelerator experiments.
The dilaton has a clear decay signal, but the production rate should 
be very small due to the extreme weak coupling.
So one need a huge luminosity to produce the dilaton from the 
accelerators.
There are, of course, other (indirect) ways to test the existence 
of the dilaton. For example, it may be worth to look for the impacts
of the dilaton in the stellar evolution and the supernovae explosion.
We will discuss these in a separate paper\cite{cho-keum}.

\vskip1.0cm
\noindent{\bf Discussion}
\vskip0.7cm

In this paper we have discussed the possible impacts of 
the hypothetical dilaton in physics, in particular in cosmology.
Remarkably the dilaton allows some definite predictions
which could be tested by experiments.
In particular, the dilaton with mass 0.5 keV or 270 MeV  
could make an exellent candidate of a dark matter.
Of course the exact value of the mass may change later because
the above results are based on the simplest assumptions.
But the importance of our analysis is that the dilaton {\it must}
be taken seriously because it exists in all the existing unified theories,
including the superstring theory.
In fact it is one of the very few predictions that 
the present unified theories can provide.
So by testing the dilaton experimentally one could test 
the unification scheme itself.
Indeed any negative experimental result on the dilaton should
make a serious damage to the credibility of the existing unified theories.
For this reason it is very important to perform experiments
which could test the existence of the dilaton.

From the theoretical point of view the importance of the above analysis
is that one can test the equivalence principle, the fifth force,
the inflation, and the dark matter problem within a single unified
picture through the dilaton.
The unified theories demand this.  
To be sure 
we still do not know whether the above unified picture is correct or not.
Nevertheless it is nice to see that the present unified theories is
able to provide such a unified picture of Nature.

\vskip1.0cm
\noindent{\bf Acknowledgement} 
\vskip0.7cm

The work is supported in part by the Korean Science and Engineering Foundation
through the Center for Theoretical Physics (SNU) and by the Ministry of
Education through the Basic Science Research Program (BSRI 97-2418).

 
\newpage
\vspace{20mm}
\begin{center}
{\bf \large Figure Captions}
\end{center}

\begin{enumerate}
\item
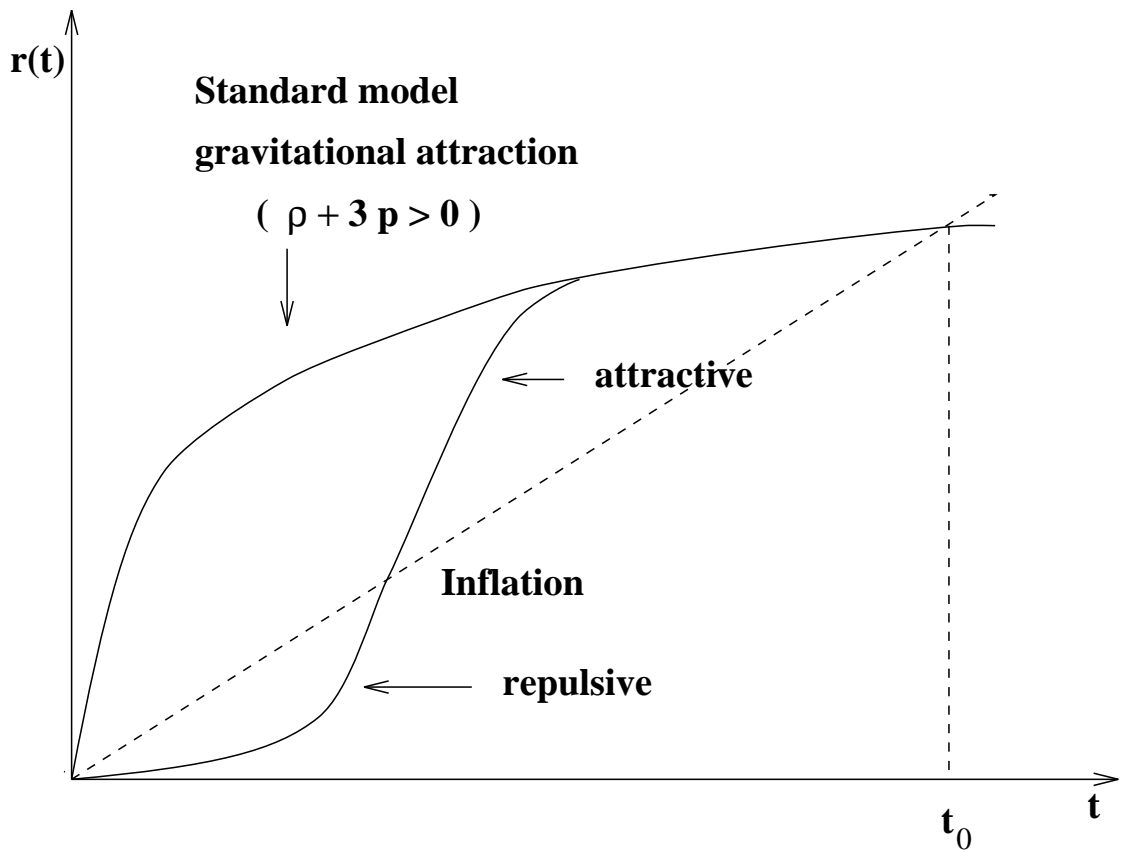
Fig. 1. A comparision between the standard cosmology
and the inflationary cosmology.
\item
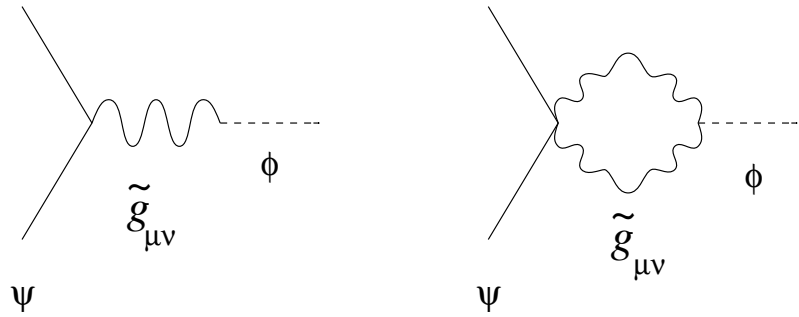
Fig. 2. The induced couplings of the dilaton to the ordinary matter 
in Brans-Dicke theory.
\item
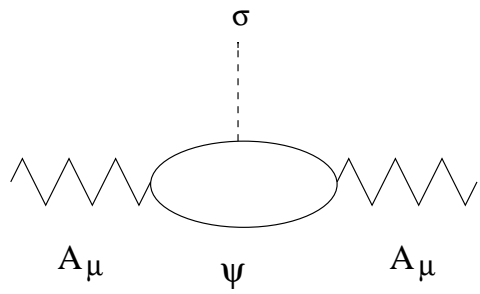
Fig. 3. An induced coupling of the dilaton to the photon.
\item
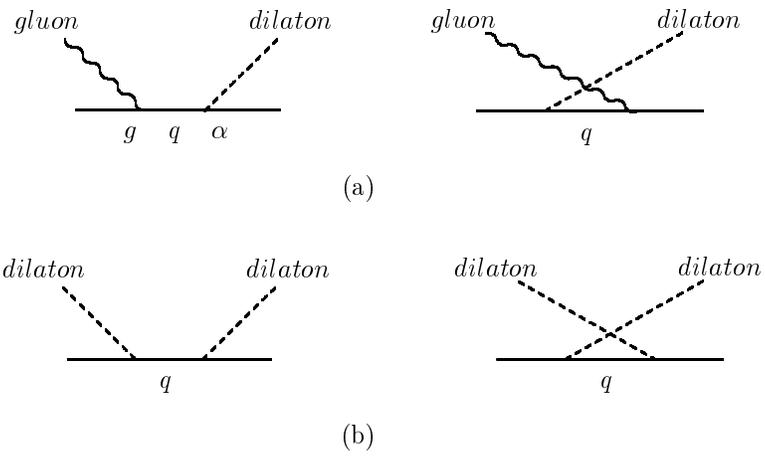
Fig. 4. The dilaton creation and annihilation process (a), and
the dilaton scattering process (b).
\item
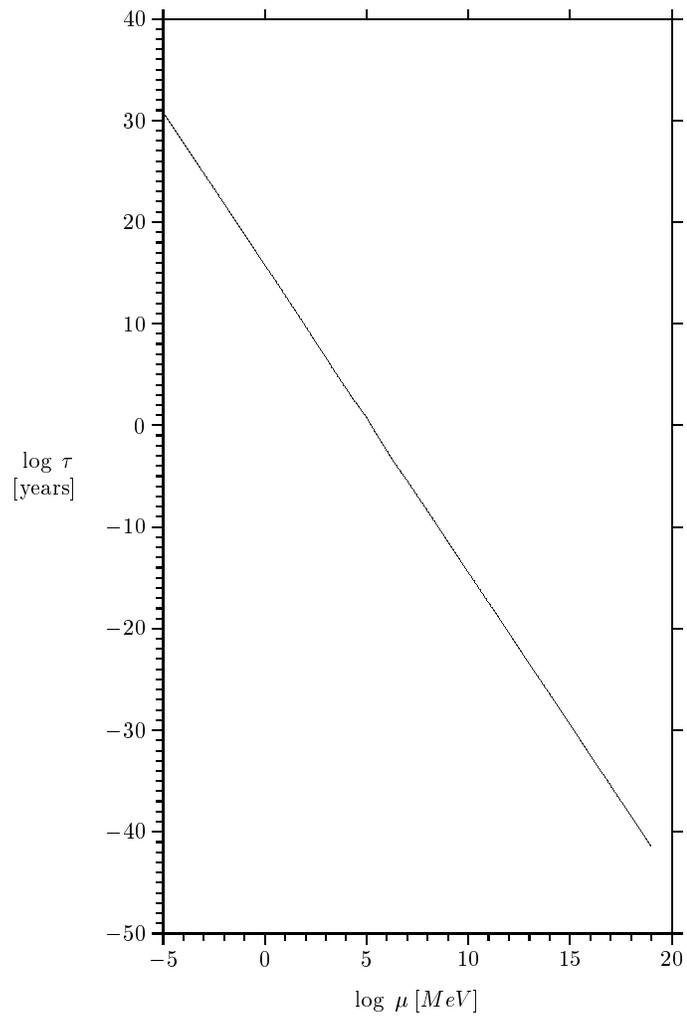
Fig. 5. The dilaton life-time versus mass based on the standard model.
\end{enumerate}

\newpage
\begin{table}[b]
\caption{
The dilatonic dark matter and its mass, decay widths, and total 
life-time for  different values of $h$.}
\begin{center}
\begin{tabular}{|c|c|c|c|c|c|}  \hline \hline
   $h$ & 0.4 & 0.5 & 0.6 & 0.7 & 0.8 
\\   \hline \hline
 $\mu$ (keV)    & 0.224 & 0.350  & 0.504 & 0.686 & 0.896  \\
 $\tau_{tot}$ ($10^{26}$ years) 
                  & 44.2& 11.6 & 3.88 & 1.54 & 0.69  \\
\hline \hline
 $\mu$ (MeV)    & 274.8 &  272.7 & 271
.0 & 269.6 & 268.3  \\
 $\Gamma_{\gamma\gamma}$ ($10^{-40}$ MeV) 
                  &  87.15 & 85.15 & 83.57 & 82.26 & 81.11 \\
 $\Gamma_{e^{+}e^{-}}$ ($10^{-55}$ MeV) 
                  &  9.60 & 9.53 & 9.47 & 9.42 & 9.38 \\
 $\Gamma_{\mu^{+}\mu^{-}}$ ($10^{-40}$ MeV) 
                  & 107.71 & 103.36 & 99.74 & 96.78 & 94.17  \\
 $\Gamma_{tot}$ ($10^{-40}$ MeV) 
                  &  195.86 & 188.55 & 183.30 & 179.04 & 175.28  \\
 $\tau_{tot}$ ($10^{10}$ years) 
                  & 0.107 & 0.111 & 0.114 & 0.116 & 0.119  \\
\hline \hline
\end{tabular}
\end{center}
\end{table}


\end{document}